%Paper: hep-th/9308139
%From: susskind@dormouse.stanford.edu (Leonard Susskind)
%Date: Fri, 27 Aug 93 16:08:29 GMT+0100
%Date (revised): Fri, 27 Aug 93 16:48:48 GMT+0100
%Date (revised): Fri, 27 Aug 93 17:05:14 GMT+0100

\input phyzzx
%\draft
\overfullrule=0pt
\hsize=6.3truein
\vsize=9.0truein
\voffset=-0.1truein

\hyphenation{Schwarz-schild}

\rightline{SU-ITP-93-21}
\rightline{August 1993}
\rightline{hep-th/9308139}

\bigskip\bigskip
\title{Strings, Black Holes and Lorentz Contraction}

\vfill
\author{Leonard Susskind\foot{susskind@dormouse.stanford.edu}}
\address{ Department of Physics \break Stanford University \break
            Stanford, CA 94305}
\vfill

\abstract
\singlespace

Consistency of quantum mechanics in black hole physics requires
unusual Lorentz transformation properties of the size and shape of
physical systems with momentum beyond the Planck scale.  A simple
parton model illustrates the kind of behavior which is needed.  It is
then shown that conventional fundamental string theory shares these
features.

\vfill\endpage

\REF\thooft{G.~`t Hooft
\journal Nucl. Phys. &B335 (90) 138}

\REF\shbhc{L.~Susskind, L.~Thorlacius, J.~Uglum, {\it The Stretched
Horizon and Black Hole Complementarity,} Stanford University
preprint, SU-ITP-93-15, hepth/9306069, June 1993, to appear in Phys.
Rev. D. \hfill\break
L.~Susskind, L.~Thorlacius, {\it Gedanken Experiments Involving Black
Holes,} Stanford University preprint,  SU-ITP-93-19, hepth/9308100
\hfill\break
Very similar ideas have been independently advocated by E. and H.
Verlinde. (H. Verlinde, private communication.)}

\REF\stpbhc{L.~Susskind, {\it String Theory and the Principle of
Black Hole Complementarity,} Stanford University preprint,
SU-ITP-93-18, hepth/9307168}

\REF\kgtssknd{J.~Kogut and L.~Susskind
\journal Phys. Rep. &8C (73) 75}

\REF\bjorken{I first heard of this breakdown of Lorentz contraction
from J. D. Bjorken. (Private communication.)}

\REF\atkwtn{J.~Atick and E.~Witten
\journal Nuclear Physics &B310 (88) 291-334}

\REF\kks{M.~Karliner, I.~Klebanov and L.~Susskind
\journal Int. Jour. Mod. Phys. &A3 (88) 1981}

\REF\parisi{J.~Kogut and L.~Susskind
\journal Phys. Rev. &D9 (74) 697 \hfill\break
A.~Casher, J.~Kogut and L.~Susskind
\journal Phys. Rev. &D9 (74) 706}

\chapter{Introduction}

G. `t Hooft has argued that the consistency of quantum mechanics in
black hole evaporation will constrain high energy physics so much
that it will determine most of its features [\thooft].  In this paper
I will show that black hole complementarity [\shbhc] implies a
radical revision of the usual kinematics of systems with very high
energy.  In particular the usual Lorentz contraction of particles
must saturate when their momenta approach the Planck scale.  In other
words the physically measurable longitudinal size of a particle must
tend to a constant and not decrease like its inverse momentum.
Furthermore the transverse size of boosted objects must grow with
momentum.  These requirements would certainly seem unbelievable if it
were not for one circumstance.  They are found to be true for the
propagation of relativistic strings.

The plan of this paper is as follows:  In this section I will make
some preliminary remarks about Lorentz boosts and the behavior of the
physically measurable dimensions of systems.  Then I will define some
concepts which will be useful in discussing the spatial localization
of information.

In section II some parton models illustrating possible behaviors of
boosted particles are presented.  One of the models is especially
interesting because it necessarily contains a massless graviton.

Section III reviews the principle of black hole complementarity and
the concept of the thermally excited stretched horizon.  We describe
how complementarity requires information, nearing the horizon, to
spread when viewed by an observer outside the black hole.  The same
information does not spatially spread when viewed by an observer in
freefall alongside the particle.  The difference between these
perceptions of events trace back to the differences in ``resolving
time'' of the apparatuses available to the two observers.  The
requirements of black hole complementarity are satisfied in the model
of section II which requires the existence of gravitons.

In section IV it is shown that string theory has exactly the
properties required by black hole complementarity [\stpbhc].  While
this does not prove that string theory is the only possible
description which allows black holes to be consistent with quantum
mechanics, it is very suggestive.

Finally in section V, I discuss the conclusions and philosophical
implications of the paper.

In classical field theory an object is described by giving the values
of certain local fields which are assumed to transform as tensors
under Lorentz transformations.  It follows straight forwardly that if
the contours of constant field strength (scalars) form spatial
spheres at rest in one frame then in a boosted frame they form
ellipsoids.  The transverse size of the ellipsoid is unchanged by the
boost and its longitudinal size is contracted according to the famous
formula of Lorentz and Fitzgerald.  In conventional quantum field
theory the situation is more complicated for a number of reasons
including the uncertainty between position and velocity and the
inability to localize a particle within its Compton wave length.
Quantum gravity may introduce other complications of an unknown kind.
 Therefore I am going to introduce a definition of size by
operational procedures which could, in principle, be used to measure
it.  We will primarily be interested in objects moving with velocity
near the speed of light.

Consider an object moving along the $z$ axis with a velocity $v
\approx 1$.  We also consider an apparatus at rest which consists of
an idealized surface of sensitive detectors such as a fluorescent
screen or a photographic plate.  I assume the grain size and spacing
is much smaller than the object.  When the object strikes the plate
it leaves a mark.  By the transverse size of the object I will mean
the size of the mark that is left.  Defined in this way it is not
clear that the transverse size remains constant as $v \rightarrow 1$.
 For example, it is widely believed that the cross section for
proton-proton scattering logarithmically increases with energy.  The
damage left by a high energy proton on a plate would also grow.  The
only requirement of Lorentz invariance is that if both the object and
the apparatus are boosted by a common angle, the transverse size not
change.  Similar conclusions can be drawn about the longitudinal size
of objects.

I believe there is a sense in which the transverse size of an object
does approach a limit as $v \rightarrow 1$ in ordinary relativistic
field theory.  Consider the difference between the spots left by
protons and neutrons.  To be more exact consider a large number of
marks left by protons and a similar ensemble left by neutrons under
otherwise identical circumstances.  Careful examination of the two
ensembles will reveal differences in the statistical properties of
p-marks and n-marks.  In ordinary QFT we do not have to study the
entire area occupied by each mark in order to distinguish the
ensembles.  Even assuming the mark size grows as $v \rightarrow 1$
the region which contains the relevant distinctions between particle
types does not.  Furthermore the longitudinal size occupied by these
distinctions, Lorentz contracts although the full physical extension
may not.

To give an idea of what I mean by the transverse size occupied by
information, consider a conventional field theory.  Assume that the
cutoff length is much smaller than the size of the objects being
studied.  Now consider two orthogonal states, $\ket{A}$ and
$\ket{B}$, which I will call particles but which could be more
general.  Let us suppose their transverse center of masses are
localized at the same place.  Now partition space into two regions.
The first region, \uppercase\expandafter{\romannumeral1}, is an
infinite solid cylinder of radius $R$ located at the same transverse
position as the centers of mass of $\ket{A}$ and $\ket{B}$.  The
second region, \uppercase\expandafter{\romannumeral2}, is its
complement.

The statistical results of all measurements within the region
\uppercase\expandafter{\romannumeral1} can be described by a density
matrix in which the degrees of freedom in
\uppercase\expandafter{\romannumeral2} are traced over.  Thus we
define
$$
\rho_A^{\rm I} \; = \;
Tr^{\rm II} \; \ket{A}\bra{A}
\eqn\one
$$

$$
\rho_B^{\rm I} \; = \;
Tr^{\rm II} \; \ket{B}\bra{B}
\eqn\two
$$
where $Tr^{\rm II}$ indicates a sum over a complete basis of states
in II.  A measure of the orthogonality of $\rho_A$ and $\rho_B$ is
defined by
$$
D_R^{\rm I}(A,B) \; = \;
{Tr \rho_A^{\rm I} \rho_B^{\rm I} \over \lbrack Tr (\rho_A^{\rm I})^2
(\rho_B^{\rm I})^2 \rbrack^{{1 \over 2}}} \;.
\eqn\three
$$

As $R \to \infty$, \ $D_R^{\rm I}(A,B)$ must tend to be zero,
indicating that the two states are orthogonal and fully
distinguishable.  Furthermore as $R \to 0$, \ $D_R^{\rm I}$ will tend
to unity since the ultraviolet behavior of all states must be the
same as the vacuum.  We can therefore define a radius $R(A,B)$ which
characterizes the transverse region in which the information
distinguishing $A$ and $B$ is localized.  For example, $R_{AB}$ could
be defined by setting $D_R (A,B)$ equal to ${1 \over 2}$.

Similarly, given a collection of states, $A, \; B, \; C$...  we can
ask for the smallest region that needs to be investigated in order to
distinguish these states.  A simple definition would be obtained by
requiring the largest of the quantities $D_R^{\rm I}(A,B)$, $D_R^{\rm
I}(A,C)$, $D_R^{\rm I}(B,C)$...  to be ${1 \over 2}$.

We can also use such density matrices to define the size of an
object.  Consider the density matrix of the outer region II when a
particle $A$ is present and when the state is pure vacuum.  Define
them by
$$
\rho_A^{\rm II} \; = \;
Tr^{\rm I} \ket{A}\bra{A}
\eqn\four
$$
$$
\rho_0^{\rm II} \; = \;
Tr^{\rm I} \ket{0}\bra{0} \;.
\eqn\five
$$

The quantity
$$
D_R^{\rm II}(A,0) \; = \;
{Tr \rho_A^{\rm II} \rho_0^{\rm II} \over \lbrack Tr (\rho_A^{\rm
II})^2
Tr (\rho_0^{\rm II})^2 \rbrack^{{1 \over 2}}} \;,
\eqn\six
$$
measures the distinguishability of the vacuum and particle $A$ in the
outer region II.  The size of $A$ can be defined by requiring
$$
D_R^{\rm II}(A,0) \; = \; {1 \over 2} \;.
\eqn\seven
$$

Generally there is no reason why the size of the particles should be
the same as the size of the regions carrying the information
distinguishing particles.  For example, if all particles in a certain
class had some sort of long range field with equal strength then the
distinction between particle types would be localized well within
their full sizes.

\chapter{Parton Models}

To understand how an observer outside a black hole describes the
behavior of matter near the horizon it is essential to first
understand ordinary field theory in the light cone frame [\kgtssknd].
 Let us introduce cartesian coordinates $(x,y,z,t)$ into flat
Minkowski space.  The $x,y$ directions will be called the transverse
directions and indicated by ($X_{\perp}$).  The combinations $\tau =
t-z$ and $X^+ = t+z$ are called light cone time and the longitudinal
direction respectively.  The conjugate to $\tau$ and $X^+$ are called
the light cone Hamiltonain and longitudinal momenta $H,P$.  For a
free particle the light cone Hamiltonian is
$$
H \; = \; {q_{\perp}^2 + m^2 \over 2P} \;,
\eqn\eight
$$
(Note that $P$ is always positive)
where $m$ is the particle's rest mass and $q_{\perp}$ is the
transverse momentum.

Light cone physics can also be thought of as the limiting description
of matter which has been boosted to very large momentum.

The space of states of light cone field theory is the Fock space
describing particles with transverse position $X_{\perp}$ and
longitudinal momentum $P$.  The states are generated by applying
creation operators $a^+(X_{\perp},P)$ on a vacuum $\ket{0}$ which is
annihilated by $a^-(X_{\perp},P)$.

Notice that quanta can have large energy either because $q_{\perp}$
is large or because $P$ is small.  For the moment let us ignore the
possibility that $q_{\perp}$ is large.  Assume that fluctuations in
transverse momenta and the mass $m$ are of some common order of
magnitude that characterizes the theory.

Let us suppose we are not interested in, or can not resolve processes
taking place on, time scales shorter than a {\it resolution time}
$\delta \tau = \epsilon$.  It is then appropriate to integrate out
all degrees of freedom with energy greater than ${1 \over \epsilon}$.
 According to \eight\ this means we integrate out quanta with
$$
P \; < \; m^2\epsilon \;.
\eqn\nine
$$
The effective description has no quanta of longitudinal momenta less
than $m^2\epsilon$.  Furthermore, in the description of a system with
total longitudinal momentum $P_{TOT}$ there can be no quanta with $P
> P_{TOT}$.

Under a longitudinal boost the P-value of each parton rescales by a
common amount.  For example, a Lorentz boost which doubles $P_{TOT}$
also doubles each constituent $P$.  However, if the resolving time
$\epsilon$ is kept fixed the lower cutoff in $P$ is not doubled.
This means that in the boosted system there will be no partons in the
allowed region $m^2 \epsilon < P < 2m^2 \epsilon$.  The partons in
the region $P \sim m^2 \epsilon$ must be dealt with separately from
the rest when a boost is performed.  They can be thought of as new
partons which come into existence solely by virtue of boosting the
system.  Feynman called these the ``wee partons''.

In certain very well behaved and uninteresting field theories the
parton distribution rapidly diminishes toward low $P$ once $P_{TOT}
>> m$.  In that case essentially no new partons are created by
increasing $P_{TOT}$.  In these theories the boost properties of
objects are very conventional.

In more interesting theories such as QCD the parton distribution is
singular near $P = 0$.  In these cases, boosting a system requires
adding partons at low $P$.  If those partons are located at ever
increasing transverse distance then the transverse size appears to
grow with $P_{TOT}$.

Similarly the longitudinal spread of the object will fail to Lorentz
contract because the constantly renewed partons are of low momentum
[\bjorken].  Nevertheless there is a sense in which particle
properties behave conventionally under boosts.  The size and shape of
the regions which contain the information necessary to distinguish
particles undergoes Lorentz contraction and no transverse spread.
This is because the distinctions are carried by the high momentum
partons which carry finite fractions of the total momentum.  The low
momentum cloud is universal.

The transverse size of an object depends both on its longitudinal
momentum and the resolution time.  However Lorentz invariance
requires it to depend only on the combination ${P_{TOT} \over
\epsilon} \;.$

We have ignored effects having to do with large transverse momenta.
These effects are interesting but do not lead to further momentum
dependence in the size of objects.  Instead they introduce fine
detail in the structure of the partons themselves [\parisi].

As we shall see, quantum gravity requires an altogether different
description when the momenta of particles begins to exceed the Planck
mass.  The new type of behavior can be illustrated by a simple model.
 Let us suppose that a particle with longitudinal momentum $P$ can be
described as a bound state of two quanta when the resolution time
$\delta \tau$ is of order $P$ in some natural units.  For simplicity
the quanta can be assumed to have approximately equal longitudinal
momenta and a transverse separation of order unity.  If the parent
particle has longitudinal momentum $P$ the constituents have ${P
\over 2}$.  The configuration is described by a wave function
$$
\psi \; = \; \psi(X_1 - X_2) \; \delta(P_1-{P \over 2}) \; \delta(P_2
- {P \over 2}) \; \delta(X_1+X_2) \;,
\eqn\ten
$$
where $X_1$ and $X_2$ represent the transverse locations of the
constituents.

Suppose when the resolution time is decreased by a factor of 2, each
constituent is itself resolved into a pair of new constituents with
the same wave functions $\psi$ except that the constituents now have
longitudinal momentum ${P \over 4}$,
$$\eqalign{
\psi(y_1,y_2,&y_3,y_4,Q_1,Q_2,Q_3,Q_4)
=  \psi({y_1{+}y_2 \over 2}  -  {y_3{+}y_4 \over 2})
\, \psi(y_1{-}y_2) \, \psi(y_3{-}y_4) \cr
&\qquad\times \delta(Q_1{-}{P \over 4}) \,
\delta(Q_2{-}{P \over 4}) \,
\delta(Q_3{-}{P \over 4}) \,
\delta(Q_4{-}{P \over 4}) \,
\delta(y_1{+}y_2{+}y_3{+}y_4) \;.
\cr}
\eqn\eleven
$$

The first factor represents the original wave function [\ten] with
$X_1$ and $X_2$ replaced by ${y_1+y_2 \over 2}$ , ${y_3+y_4 \over 2}$
respectively.  The second and third factors represent the
compositeness of the original constituents.

Let us continue this process so that each time we improve the
resolution by a factor of 2.  The previous constituents are resolved
into pairs with the wave function $\psi$.  After $n$ iterations the
resolving time is
$$
\epsilon \sim P({1 \over 2})^n \;,
\eqn\twelve
$$
the number of constituents is $2^n$, and the longitudinal momentum of
each is ${P \over 2^n}$.

As the resolving time decreases the transverse spread of the
configuration tends to a gaussian probability distribution for
finding a constituent at a given transverse distance.  The width of
the gaussian grows like the square root of the number of iterations.
Thus the transverse size $R$ is given by
$$
R({P\over\epsilon})\sim\sqrt{\ln{{R\over\epsilon}}} \;.
\eqn\thirteen
$$

This formula describes a growth which is similar to that of the
proton that I described before.  However this time the information is
spread over the entire area.  To see this we can consider
constructing a second state, replacing \ten\ by
$$
\psi' \; = \; \psi'(X_1-X_2) \;
\delta(P_1-P_2) \;
\delta(P_2-{P \over 2}) \;
\delta(X_1+X_2) \;,
\eqn\fourteen
$$
where $\psi'$ is orthogonal to $\psi$.  In each iteration we still
replace each constituent by a pair with the original wave function
$\psi$.  After any number of iterations the two wave functions are
orthogonal.  However the density matrices for bounded regions of
fixed size $R_0$ are indistinguishable as ${P \over \epsilon} \to
\infty$.  To detect the orthogonality of the two states a region of
size $R({P \over \epsilon}) \sim \ln{{P \over \epsilon}}$ must be
inspected.  As we shall see this is a fundamental property that
quantum gravity must have if black hole evaporation is to be
consistent with quantum mechanics.

The longitudinal size of the distribution can also be estimated.
Since the individual constituent longitudinal momenta are of order
$\epsilon$, the resolving time, the uncertainty principle suggests
that the longitudinal size $\Delta z$ satisfies $\Delta z \sim {1
\over \epsilon}$.  In conventional terms this is equivalent to an
absence of longitudinal Lorentz contraction as $ P \to \infty$ with
fixed $\epsilon$.

I will now argue that if such a model can be consistent with special
relativity it must contain a graviton.  To see this let us consider
the scattering of two particles.  We take one particle to be at rest
and one moving along the z-axis with large momentum $P$.  The fast
particle has longitudinal momentum $P$ and the scattering can resolve
internal motions with $\delta \tau \approx 1$ so that the fast
particle must be described as a number $N \sim P$ constituents.  Now
consider the low momentum transfer elastic amplitude.  Let $q$ be the
transverse momentum transfer.  Since the target can scatter off any
of the constituents the amplitude will be proportional to $N$.
Furthermore, since the spatial distribution of constituents is
gaussian with width of order $(\ln{N})^{1 \over 2}$ we find the
amplitude to be
$$A(q^2) \; \sim \; e^{-(\ln{N})q^2}N \; \sim \; P^{1-q^2} \;.
\eqn\fifteen
$$

The reader will recognize this as a Regge behaved scattering
amplitude corresponding to a linear Regge trajectory
$$
J(q^2) \; = \; 2 \; - \; q^2
$$
from which one deduces the existence of a massless spin 2 particle.

The above argument is not meant to be a serious mathematical proof.
It is a paraphrasing of string theory which we will see has the
properties of the model.  The main features to remember in this model
are that the spatial extension of the cloud of information carried by
a particle has longitudinal and transverse extension which depends on
the ratio of the longitudinal momentum and resolution time.  The
pattern of transverse growth that occurs as the resolution time is
decreased is similar to a common feature of many systems known as
{\it branching diffusion}.

\chapter{Implications of Black Hole Complementarity}

Consider an object falling toward the horizon of a black hole.  From
the viewpoint of fiducial observers at fixed static position, the
momentum of the object increases without bound and its internal
motions slow indefinitely.  In effect, the fiducial observers outside
the black hole see the object with increasing powers of resolution.
To follow this process into the stretched horizon at a few Planck
lengths from the event horizon the Lorentz boost properties of matter
must be thoroughly understood.

The black hole hole can be described by external observers in terms
of tortoise coordinates which cover only the exterior region.
Tortoise time is identical to Schwarzschild time and the tortoise
radical coordinate $r^*$ is defined by
$$
r^* \; = \; r \; + \; 2m \log{(r-2m)}\;.
\eqn\rstar
$$
Far from the horizon the metric has the flat space form
$$
ds^2 \; = \; dt^2 \;-(dr^*)^2 \; -(r^*)^2d\Omega^2 \;.
\eqn\fspace
$$
Near the horizon it locally behaves like
$$
ds^2 \; = \; \big ({e^{{r^* \over 2m}} \over 2m}\big ) \; \lbrack
-(dr^*)^2 \; + \; dt^2 \rbrack -dX^2_{\perp} \;,
\eqn\behavior
$$
where $X_{\perp}$ are cartesian coordinates transverse to the radial
direction.

As the particle falls toward the horizon its longitudinal momentum
increases like $\exp{({t \over 4m})}$.  If the system behaves like a
conventional classical object it will appear to have fixed transverse
size and Lorentz contracted longitudinal extension.  The center of
the object will move on a trajectory which approaches
$$
r^* \; + \; t \; = \; 0 \;,
\eqn\trajectory
$$
as $t \to \infty $ and its longitudinal extension $\Delta r^*$ will
satisfy
$$
e^{{r^* \over 4m}} \Delta r^* \; \approx \; e^{-t \over 4m} \;,
\eqn\satsfy
$$
or
$$
\Delta r^* \; \sim \; 1 \;.
\eqn\short
$$
Eventually the particle and all its structure disappears to $r^* =
-\infty$ and is lost.  At best the information can be retrieved at
the very end of the Hawking evaporation.  It is this picture that is
implicit when conventional quantum field theory is studied in curved
space-time.

Black hole complementarity requires a different behavior from the
viewpoint of the external observer.  The information carried by the
object should get deposited in a layer called the stretched horizon
which is located near $r^* = 0$.  This is the layer where the local
temperature of the Unruh radiation is of Planckian magnitude.
Furthermore, the information is supposed to be distributed among the
hypothetical stretched horizon degrees of freedom as if it was being
thermalized.  In the final state of thermal equilibrium the
information should be delocalized over the entire horizon.  A
reasonable guess is that the information diffuses so that its
transverse spread grows like
$$
R^2 \; \sim \; {t\over 4M} \;,
\eqn\grow
$$
we use ${t\over 4M}$ because the proper time on the stretched horizon
is red shifted relative to Schwarzschild time by a factor of order
$4M$.

The longitudinal spread of the information implied by complementarity
is also unconventional.  The region occupied by the system must
continue to overlap the stretched horizon near $r^* = 0$.  This
requires \satsfy\ and \short\ to be replaced by
$$
\Delta r^* \; \sim \; e^{{t \over 4M}} \;.
\eqn\replace
$$
This is equivalent to the condition that no Lorentz contraction takes
place once the particles falling into the black hole reach momenta of
order the Planck mass.  In other words the longitudinal extension
$\Delta z$ should satisfy
$$
\vert \Delta z \vert \; \sim \; 1 \;,
\eqn\longitude
$$
in Planck units.

The conditions \replace\ and \longitude\ are just those satisfied by
the branched diffusion model of Section III.  Of course the model was
cooked up for just this purpose.  However it is interesting that it
also leads to the existence of a massless spin two graviton.

\chapter{Strings Near a Horizon}

String propagation in a Schwarzschild background has not been
completely
analyzed.  However, the region near the event horizon of a very large
black
hole is enough like Minkowski space that much of what we need to know
can be
determined.  In particular, as long as the region under study is
small in
all its dimensions compared with the Schwarzschild radius, the
external
region of the horizon is isomorphic to Rindler space.

In the previous sections we assumed that the standard laws of physics
hold
down to the Planck scale.  In string theory however, the new physics
begins
at the {\it string scale} which differs from the Planck scale by
factors of
the dimensionless coupling constant $g$.  If $\ell_p$ is the Planck
length and
$\ell_s$ is the string length then
$$
\ell_p \; = \; g \ell_s \;,
\eqn\slength
$$
so that if $g$ is very small the new physics begins at length scales
appreciably larger than $\ell_p$.  In this case the stretched horizon
should
be placed a distance $\sim \ell_s$ from the event horizon.  At this
distance
the Unruh temperature is the same as the Hagadorn temperature and the
properties of the vacuum become markedly different [\atkwtn] from the
zero temperature vacuum seen by a freely falling Minkowski observer.
This is
also the place where the standard rules of Lorentz contraction begin
to fail
and where information begins to transversely spread [\stpbhc].  The
Planck length is where perturbative string theory fails but the
interesting
physics seen by the outside observers takes place between $\ell_s$
and
$\ell_p$.

Let us compute the properties of free strings in the light cone
frame.
Points of the string are described by a transverse location
$X_{\perp}(\sigma)$ and whatever internal degrees of freedom are
implied by
supersymmetry and compactification.  In  the light cone frame the
internal
degrees of freedom are decoupled from $X_{\perp}(\sigma)$.  Thus the
normal
mode decomposition of $X_{\perp}$ is the same as for free bosonic
strings
$$
X_{\perp}(\sigma) \; = \; X_{\perp}(C.M.) \; + \;
\sum_{\ell} {X(\ell) \over \ell}
e^{i \ell \sigma} \; + \;
{\widetilde X (\ell) \over \ell} e^{-i \ell \sigma} \;.
\eqn\boson
$$
The transverse size of the string can be estimated by computing the
quantity
$$
R_{\perp}^2 \; = \;  \langle\lbrack X_{\perp}
(\sigma) - X_{\perp} (c.m.) \rbrack^2 \rangle \;,
\eqn\tsize
$$
where the expectation value is calculated in whatever state is under
consideration.  For the ground state one easily finds
$$
R_{\perp}^2 \; = \; \sum_{\ell} \; {1\over \ell} \;,
\eqn\gstate
$$
which diverges logarithmically.  This divergence is the key to
understanding
the curious properties of strings under Lorentz boosts.

The divergence  in $R_{\perp}^2$ is due to a summation over modes of
arbitrarily high frequency.  The frequency of the $\ell^{{\underline
{th}
}}$  normal mode in light cone time $\tau$ is
$$
\nu_{\ell} \; = \; {\ell \over P_{TOT}}
\;,
\eqn\mode
$$
where $P_{TOT}$ is the longitudinal momentum of the string.  In
\mode\ the
frequency is defined in string units.  If an experiment is performed
by an
observer with a resolution time $\epsilon$ then the modes with $\nu >
{1
\over \epsilon}$ should be cut off.  Hence we define a resolution
dependent
size $R_{\perp}^2 (\epsilon)$ by
$$
R_{\perp}^2 (\epsilon) \; = \;
\sum_{\ell=1}^{{P_{TOT} \over \epsilon}}
{1 \over \ell} \; \approx \; \log{{P_{TOT}
\over \epsilon}} \;.
\eqn\resdsize
$$
Evidently the transverse size grows exactly as in the branched
diffusion
model of Section II [\stpbhc , \kks].  The phenomenon of transverse
growth
with decreasing resolution time has been noted previously but its
connection
with black hole horizon's has not.  The longitudinal behavior as a
function
of resolution has not, to my knowledge, received any attention.  To
compute
the mean longitudinal spread $\Delta z$ we use the constraint
equation
$$
{\partial X^+ \over \partial \sigma} \; = \;
{\partial X_{\perp} \over
\partial \sigma} \;  {\partial X_{\perp} \over \partial \tau} \; + I
\;,
\eqn\constrnt
$$
where $\ell$ represents the contribution from compactified modes,
fermionic
modes etc.  We can rewrite \constrnt\ in terms of Virasoro generators
$$
{\partial X^+ \over \partial \sigma} \; = \;
\sum_{\ell} L(\ell) \; e^{i \ell \sigma} -
\widetilde L (\ell) \; e^{-i \ell \sigma} \;,
\eqn\virasoro
$$
which can be integrated to give
$$
X^+ \; = \; X^+(c.m.) \; + \; \sum \;
{L(\ell) \; e^{i \ell \sigma} \over i \ell} \; + \;
{\widetilde L (\ell) \over i \ell} \;
e^{-i \ell \sigma} \;.
\eqn\intrte
$$
Using the standard Virasoro algebra one finds
$$
\bra{0} \Big ( X^+ (\sigma) - X^+ (c.m.) \Big )^2 \ket{0}
\; \sim \; \sum_{\ell} \ell \;,
\eqn\algebra
$$
which diverges quadratically.  The cure is as before.  Averaging
$X^+$ over a
resolution time $\epsilon$ cuts off the sum at $\ell \; \sim \; {P
\over
\epsilon}$.  Thus
$$
\bra{0} \Big ( X^+ (\sigma) - X^+ (c.m.) \Big )^2 \ket{0}
\; \sim \; {P^2 \over \epsilon^2} \;,
\eqn\algebra
$$
or
$$
\vert \Delta X^+ \vert \; \sim \; {P \over \epsilon} \;.
\eqn\endpgph
$$

Equation \endpgph\ indicates that no Lorentz contraction of the
string
distribution takes place.  The two properties \resdsize\ and
\endpgph\ are
precisely what is needed in order that an infalling string appears to
spread
over the stretched horizon without escaping to $r^* = -\infty$.

The spreading process begins to occur when the string reaches the
stretched
horizon at distance $\ell_s$ from the event horizon.  The process is
very
similar to the stochastic evolution of a scalar field in an inflating
universe.  In both cases more and more modes enter the description
with time.
These modes enter with random  phase and amplitude.  In each case the
growth
and spreading over the {\it target space} can be described by
stochastic
interactions with a heat bath.  In the string case the heat bath is
provided
by the Unruh effect.

If no other effects take place the string would grow to a size
comparable to
the Schwarzschild radius in a time given by
$$
t \; = \; g^2 M^3 \;,
\eqn\enough
$$
in Planck units.  If $g$ is small this is a short time by comparison
with the
evaporation time of the black hole.

As the string replicates its transverse density increases.  At the
center of
the distribution the average number of strings $N$ passing through a
region of
area $A$ (measured in string units) is of order $\exp{R_{\perp}}^2
\sim e^{{t
\over 4M}}$.  However, this enormous density of string certainly
leads to new
effects once it becomes of order ${1 \over g^2}$.  At this time the
probability for string interactions becomes unity and perturbation
theory
breaks down.  One attractive possibility is that the growth of string
density
is cut off at this point.  The result would be that the density grows
until
there is about one string per unit Planck area.  This is also
suggested by the
fact that the entropy of a black hole is proportional to its area as
measured
in Planck units.

Perhaps the most remarkable aspect of the above description is that
none of it
is seen by an observer who falls through the horizon with the string.
 Such an
observer sees the string with a fixed time resolution and therefore
sees a
constant transverse and longitudinal size as the horizon is crossed.

\chapter{Philosophical Implications}

Black hole complementarity and its realization in string theory imply
profound
changes in our current views of matter and space-time.  These
concepts further
erode the classical realism of the Newtonian picture of the universe.
 They
entail a new degree of relativity and observer dependence of reality.
 The
special theory of relativity destroyed the invariant meaning of
simultaneity.
Quantum theory introduced the idea of incompatible measurements and
eliminated
the classical concept of a well defined trajectory.  What was left
intact is
the invariant event, occurring in a well defined space-time location
even if
that event can only be predicted statistically.  Now however, even
that can no
longer be relied upon.  Consider, for example, that the destruction
of an
individual falling into a black hole takes place in a space-time
region and in
a manner which appears entirely different to observers in free fall
and those
supported outside the horizon.  To those in free fall the individual
easily
survives the passage through the horizon but is destroyed by infinite
tidal
forces much later.  The outside observer witnesses death by heat at
the
stretched horizon.  Which is correct?  In my view there is no more an
answer
to this question than to whether two events really are simultaneous
or to
which of the two paths a photon traveled.

All of this is possible only because matter is not anchored in
space-time as in
classical or quantum field theory.  The more precisely one tries to
resolve
the location of the constituents of matter the more they fluctuate to
large
distances.  Probing strings with infinite time resolution reveals
that each
bit of string fills space out to infinite distances.  Only because
finite
energy implies finite time resolution do we see localized matter.

\chapter{Acknowledgements}

The author thanks A. Peet and L. Thorlacius for help in calculating
the
longitudinal spread of strings.  He also thanks A. Mezhlumian for
informative
discussions about branched diffusion processes, and J. Susskind for
technical
assistance.  The work was supported in part by NSF grant PHY 89
17438.

\refout
\doublespace

\end